\documentclass[runningheads]{llncs}
\usepackage[T1]{fontenc}
\usepackage[misc]{ifsym}
\usepackage{graphicx}
\usepackage{amssymb}
\usepackage{makecell}
\usepackage{algorithm}
\usepackage{algorithmic}
\usepackage{amsmath,amsfonts}
\usepackage{multirow}
\usepackage{diagbox}
\usepackage[symbol]{footmisc}
\usepackage{subfigure}
\usepackage[colorlinks,
            linkcolor=blue,       
            anchorcolor=blue,  
            citecolor=blue, 
            urlcolor=blue,
            ]{hyperref}

\begin{document}
\title{Towards Confidential and Efficient LLM Inference with Dual Privacy Protection}
\titlerunning{Confidential and Efficient LLM Inference}
%
\author{
Honglan Yu\inst{1,2,3} \and 
Yibin Wang\inst{1,2,3} \and 
Feifei Dai\inst{1} \and 
Dong Liu\inst{1} \and 
Haihui Fan\inst{1} \and 
Xiaoyan Gu\inst{1,2,3}\textsuperscript{(\Letter)}
}
\authorrunning{H. Yu et al.}
%
\institute{Institute of Information Engineering, Chinese Academy of Sciences, Beijing, China\\
\and
School of Cyber Security, University of Chinese Academy of Sciences, Beijing, China\\
\and 
State Key Laboratory of Cyberspace Security Defense, Beijing, China\\
\email{\{yuhonglan, wangyibin, daifeifei, liudong, fanhaihui, guxiaoyan\}@iie.ac.cn}
}

\maketitle              
\begin{abstract}
CPU-based trusted execution environments (TEEs) and differential privacy (DP) have gained wide applications for private inference. Due to high inference latency in TEEs, researchers use partition-based approaches that offload linear model components to GPUs. However, dense nonlinear layers of large language models (LLMs) result in significant communication overhead between TEEs and GPUs. DP-based approaches apply random noise to protect data privacy, but this compromises LLM performance and semantic understanding. To overcome the above drawbacks, this paper proposes CMIF, a Confidential and efficient Model Inference Framework. CMIF confidentially deploys the embedding layer in the client-side TEE and subsequent layers on GPU servers. Meanwhile, it optimizes the Report-Noisy-Max mechanism to protect sensitive inputs with a slight decrease in model performance. Extensive experiments on Llama-series models demonstrate that CMIF reduces additional inference overhead in TEEs while preserving user data privacy.

\keywords{TEE \and Differential Privacy \and Private Inference}
\end{abstract}

\section{Introduction}\label{sec:intro}

Nowadays, large language models (LLMs) are increasingly incorporated into many aspects of our life, such as autonomous driving~\cite{BEVFormer}, disease diagnosis~\cite{thirunavukarasu2023large} and machine translation~\cite{xu2024a}. LLMs are typically deployed on cloud servers. Users can access these models via websites or APIs. However, this approach raises privacy concerns, as personal data, such as symptom-related queries, can be collected by servers or providers for commercial use. Therefore, it is necessary to protect data privacy during model inference.

As TEE systems gradually develop, various TEE-based privacy protection mechanisms have emerged. TEEs, secure CPU areas, offer hardware-level protection for internal code and data~\cite{CML2021}. To balance data privacy and inference efficiency, researchers partition deep learning models to run on TEEs and untrusted GPUs. Most TEE-based methods adopt a partition-after-training approach~\cite{ShadowNet,xu2024tempo,slalom2019iclr,SOTER,2021Darknight}. 
They encrypt linear layers via linear transformations for accelerators while keeping nonlinear layers, like activation layers, within TEEs for confidentiality.  
Recently, the training-after-partition paradigm has been proposed in TEESlice~\cite{Slice2024}. They separate and train privacy-related layers from pre-trained models. Furthermore,~\cite{huang2024fast} combine TEESlice with parameter-efficient fine-tuning methods in LLMs for precise private parameter protection. For user data, many TEE-based approaches employ One Time Pad (OTP)~\cite{slalom2019iclr} encryption, which generates random vectors to encrypt representations.

Differential privacy (DP) is another mechanism, used to enhance data privacy in LLMs during fine-tuning and inference. DP-Forward~\cite{DP-Forward} and TextObfuscator~\cite{zhou-etal-2023-textobfuscator} propose perturbing embeddings with calibrated noise. These methods obfuscate original word representations, making it challenging for attackers to recover sensitive information. Meanwhile, SANTEXT+~\cite{SANTEXT} and CUSTEXT~\cite{chen-etal-2023-customized} sanitize user inputs by identifying and replacing sensitive tokens with semantically similar alternatives, enabling the sharing of sanitized text without compromising privacy.

\begin{figure}[t]
\centering
\includegraphics[scale=0.35]{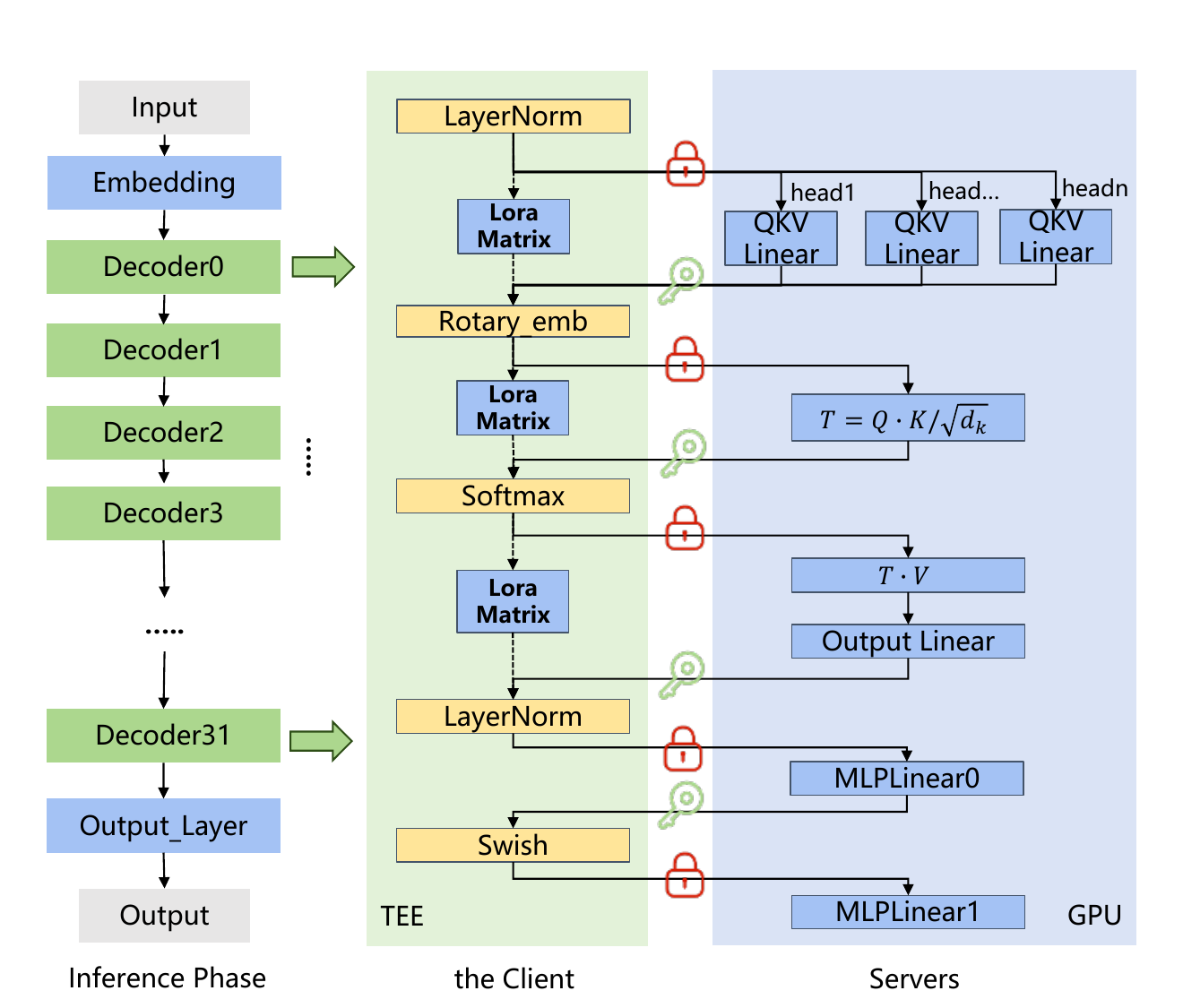} 
\caption{The TEE-based Method for Llama2-7B. Blue operations denote linear operations, while yellow ones represent nonlinear operations.}
\label{fig1}
\vspace{-0.4cm}
\end{figure}

However, despite the undeniable contributions of previous approaches, they have notable drawbacks when applied to LLMs. 
Firstly, TEE-based model partition approaches incur significant time overhead during inference due to data transmission between TEEs and GPUs~\cite{huang2024fast}. As illustrated in Fig.~\ref{fig1}, existing methods deploy the most sensitive parts in TEEs and use OTP to encrypt features transmitted between GPUs and TEEs. Since OTP only supports linear operations, the nonlinear layers of the LLM must stay inside TEEs, leading to frequent switching and increased inference latency. 
Secondly, applying existing DP-based methods to LLMs proves challenging, as random perturbations introduced by DP  can disrupt the semantic understanding of LLMs, resulting in incoherent and incorrect outputs.
Furthermore, these DP-based methods require access to LLM parameters and architectures for privacy protection, potentially compromising the security of the models. 

To address these issues, we propose CMIF, a Confidential and efficient Model Inference Framework. Our motivation is to protect user data privacy through a sanitization mechanism for input tokens. It replaces sensitive words while preserving semantic coherence, so that the model does not need to return TEE for decryption during inference, reducing the communication and overhead associated with linear encryption mechanisms. 
The semantic-based sanitization employs the Report-Noisy-Max algorithm to protect user data privacy while maintaining semantic coherence. 
Simultaneously, TEEs secure both the sanitization mechanism and client-side models. The code is available at~\href{https://github.com/Grace-byte912/CMIF}{https://github.com/Grace-byte912/CMIF}. In summary, the key contributions of this paper are as follows:
\begin{enumerate}
\item We introduce CMIF, an innovative confidential and efficient inference framework for LLMs. CMIF safeguards data privacy through an advanced text sanitization mechanism and enhances inference efficiency by leveraging GPU acceleration on servers.

\item We optimize the RNM-based text sanitization mechanism, improving model performance while maintaining data privacy. Experiments on benchmark datasets show that our mechanism outperforms state-of-the-art methods at comparable privacy levels.

\item We conduct extensive evaluations of CMIF's performance, demonstrating significant improvements in inference efficiency compared to previous TEE-based methods.

\end{enumerate}

\section{Preliminaries}
\subsection{Large Language Models}
Large Language Models (LLMs) such as LlaMA2 \cite{touvron2023llama2openfoundation} contain hundreds of billions of parameters. These models are pre-trained on large, high-quality corpora, then fine-tuned to align their outputs with human preferences.
LLMs have shown effectiveness in diverse downstream tasks through text generation.
Typically, LLMs adopt the decoder-only architecture for better parallelization and scalability. They can be divided into three main parts: the embedding layer, decoder layers, and output layer, as shown in Fig.~\ref{fig1}.

\subsection{Trusted Execution Environment}
To enhance system safety, TEEs~\cite{tee2015} operate in parallel with the primary OS. TEEs establish a secure, isolated execution environment that protects the confidentiality and integrity of code and data, even if the main OS is breached. Additionally, TEEs also establish secure communication with external devices. 
Intel's Software Guard Extensions (SGX)~\cite{IntelSGX}, a common TEE implementation, offers security-focused instruction sets for Intel processors, enabling the creation of private memory regions known as enclaves.
SGX developers must partition their applications into enclave and non-enclave halves, incurring significant refactoring costs. Research~\cite{Graphene-SGX2017,occlum2020} explore using LibOSes to reduce this cost by enabling programs to run inside enclaves with minimal or no modifications. 

\subsection{Local Differential Privacy}
Differential Privacy (DP) ensures individual data privacy by adding noise to statistical query results, preventing accurate reconstruction or inference of individual information, even with multiple queries. 
Local differential privacy (LDP)~\cite{bassily2015local} allows data owners to perturb their data with a randomized algorithm locally before sending it to untrusted data collectors.


\begin{definition}[Sensitivity of Function]
For function $f(\cdot)$ and two neighboring inputs $\mathcal{X}, \mathcal{X}^{\prime}$, the sensitivity $\Delta({f})$ is defined as:
\begin{equation}
\Delta({f})=\sup_{\mathcal{X}\simeq\mathcal{X}^{\prime}}||f(\mathcal{X})-f(\mathcal{X}^{\prime})||_F,
\end{equation}
where $||\cdot||_F$ represents the Frobenius norm. 
\end{definition}

\begin{definition}[Report-Noisy-Max Algorithm]
Given a finite set of outcomes $\mathcal{Y}$ and a score function $d: (\mathcal{D}, y) \rightarrow \mathbb{R}$, for each outcome $y \in \mathcal{Y}$, 
\begin{equation}
\mathcal{M}_d(\mathcal{D})={\arg\max_{y \in \mathcal{Y}} \left(d(\mathcal{D},y) + \text{Lap}\left(\frac{\Delta}{\epsilon}\right)\right)},
\end{equation}
where $\text{Lap}(\frac{\Delta}{\epsilon})$ is the Laplace distribution with scale $\frac{\Delta}{\epsilon}$. $\epsilon$ represents the privacy budget. Users can modify $\epsilon$ to control the level of privacy protection. 
\end{definition}

\textbf{Theorem 1}\label{theorem1}: If the score function is $\Delta$-sensitive then the Report-Noisy-Max algorithm is $\epsilon$-differentially private~\cite{DPfoundations}. 



\section{Method}
\begin{figure}[t]
\centering
\includegraphics[scale=0.27]{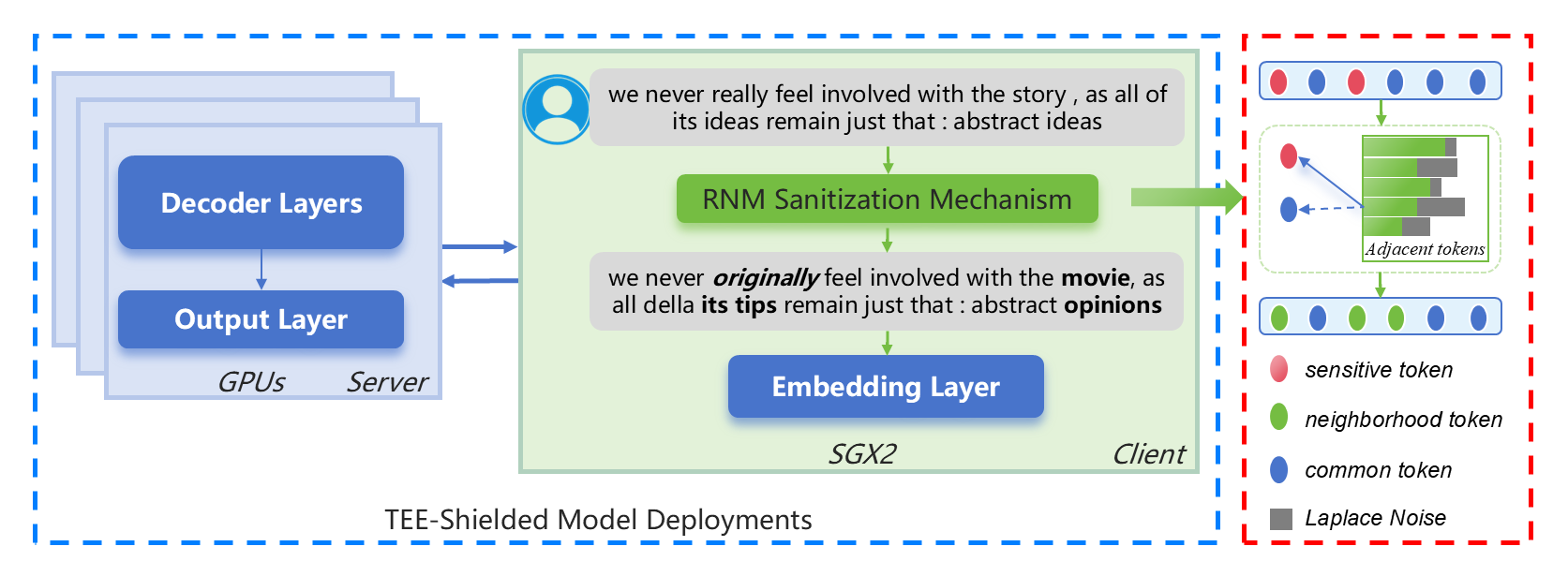} 
\caption{An overview of CMIF. Sensitive tokens are replaced by adjacent tokens based Report-Noisy-Max (RNM) sanitization mechanism (the red box).}
\label{fig2}
 \vspace{-0.4cm}
\end{figure}

As shown in Fig.~\ref{fig2}, CMIF consists of two main components: secure model deployment with TEEs, and Report-Noisy-Max (RNM) sanitization mechanism. 

\subsection{TEE-shielded Model Deployments} 
To reduce TEE-GPU communication latency and ensure security, the model is partitioned into two parts: the embedding layer, which maps input tokens to dense vectors, is deployed in secure enclaves of SGX2 on the local machine. This ensures that sensitive data is processed in a trusted environment and protect the confidentiality and integrity of client-side sub-models.
Given the substantial computational resources and financial investments required for LLMs, it is crucial to protect their parameters. 
The remaining layers, which perform the core language understanding and generation, are deployed on powerful remote servers equipped with high-performance GPUs. PyTorch's RPC mechanism efficiently facilitates communication and data transfer between these components.

\subsection{Report-Noisy-Max Sanitization Mechanism}
To mitigate performance degradation while preserving privacy with DP mechanisms, we introduce the Report-Nosiy-Max (RNM) sanitization mechanism. Following previous research~\cite{chen-etal-2023-customized,SANTEXT}, we regard low-frequency words as sensitive, which need to be protected most. Others are non-sensitive. This mechanism consists of two primary functions: a distance function and a mapping function. The distance function identifies substitute candidates for each word, and the mapping function applies the probabilistic RNM strategy to replace words with their candidates.

\textbf{Distance Function.}
For each word $w_i$ in the vocabulary $V$, we search for its top-$k$ candidate words (excluding the word itself) based on cosine similarity, which is advantageous in comparing the semantic distance between two words. By selecting the word that maximizes the cosine similarity with the target, we can identify its semantical substitute while minimizing the impact on the model's performance. 

Subsequently, the distances of words in the candidate list are normalized, ensuring that the sensitivity of the score function $cos_{norm}$ is equalized to 1. 
All distances are normalized as follows:
\begin{equation}
cos_{norm}(\mathbf{w}, \mathbf{w'}) = \frac{cos(\mathbf{w}, \mathbf{w'}) - cos_{min}}{cos_{max} - cos_{min}} = \frac{\frac{\mathbf{w_i} \cdot \mathbf{w_{i}'}}{\left | \mathbf{w_i} \right |_2 \left | \mathbf{w_{i}'} \right |_2} - cos_{min}}{cos_{max} - cos_{min}},
\label{eq:cos_norm} 
\end{equation}
where $\mathbf{w_i}$ represents the $i$th tensor in the embedding layer of LLMs and $cos_{min}$ and $cos_{max}$ denote the minimum and maximum values among the candidate list.

\textbf{Mapping Function.}
During inference, the mapping function $\mathcal{F}$ sanitizes user data before uploading it to servers. Specifically, for each word in user data, $\mathcal{F}$ first checks whether it is sensitive based on the frequency threshold.
For sensitive words, $\mathcal{F}$ replaces them with optimal substitutes selected through Report-Noisy-Max algorithm, which adds noise drawn from the Laplace distribution to each substitute's score. The noise strength is determined by the sensitivity of $cos_{norm}$ divided by privacy budget $\epsilon$. The RNM algorithm returns the word with the highest noisy score. 
For non-sensitive words, $\mathcal{F}$ adopts a probabilistic mechanism, performing the same algorithm with probability $p$. If not selected, they remain unchanged. Users can adjust $p$ as a hyper-parameter: a smaller $p$ leads to greater fidelity to the original words. 



 
\section{Experiments}

\subsection{Experimental Setup}

We develop an implementation of CMIF based on the distributed RPC framework in PyTorch and evaluate it on Occlum v1.0~\cite{occlum2020}, which enables programs to run on SGX2 with a few modifications. The implementation is evaluated on an Intel(R) Xeon(R) Gold 6326 CPU processor with 499GB RAM in the client. The server is equipped with two NVIDIA A100 GPUs (80GB) and 250GB RAM. $k$ is set to 30 and the replacement probability for non-sensitive words is 0.3 by default. 
Tokens with frequencies in the bottom 20\% are sensitive words. 

\textbf{Threat Model}
We assume that model owners and cloud providers are semi-honest. Similar to previous studies, we do not consider side-channel attacks on TEEs~\cite{ShadowNet,huang2024fast,xu2024tempo} because they are beyond the scope of this paper. The TEE provides an isolated and secure environment where programs can be executed with integrity and confidentiality. 


\textbf{Datasets and Metrics} We evaluate the performance of CMIF mainly on three typical natural language processing datasets: SST-2 and QNLI for text classification, DialogSUM for dialogue summarization, and IFEval for instruction following. 
We use different metrics for evaluations. We select precision, recall, and accuracy for SST-2 dataset. For the DialogSUM dataset, we use ROUGE-1 (R-1) score, ROUGE-2 (R-2), and ROUGE-L (R-L) score.
IFEval dataset has its automatic, rule-based metric\cite{zhou2023instructionfollowingevaluationlargelanguage}. We employ the model average forwarding time to measure its inference efficiency. 

We use three popular light LLMs, including bert-base-uncased~\cite{DBLP:BERT}, Llama2-7B-chat \cite{touvron2023llama2openfoundation}, and Llama3-8B-Instruct \cite{meta2023llama3} (denoted as Bert, Llama2-7B, and Llama3-8B) to evaluate CMIF. 

\subsection{Model Performance}
\label{modelper}
We compare the classification accuracy of different DP mechanisms based on Bert. The evaluation results are shown in Table~\ref{Table2}. As $\epsilon$ increases, the accuracy of Bert increases slightly because there are fewer perturbations caused by noise. Furthermore, our method consistently outperforms other methods under different privacy budgets. This superior performance is mainly attributed to our semantic-based sanitization mechanism. For different models, we adopt different word vectors according to parameters in the embedding layer. Both operations ensure that the selected word is semantically close to the original one. However, CUSTEXT+~\cite{chen-etal-2023-customized} adopts GloVe~\cite{pennington-etal-2014-glove} embeddings as word vectors, which can conflict with the pre-trained model embeddings.

\begin{table}[t]
\caption{\small Utility comparison of different sanitization mechanisms at similar privacy levels. For SST-2 dataset, $k$ is 30 and for QNLI dataset, $k$ is 20.}
\centering
 {
\small{
\begin{tabular}{|c|p{1.2cm}|p{1.2cm}|p{1.2cm}|p{1.2cm}|p{1.2cm}|p{1.2cm}|}
\hline
\multicolumn{1}{|c|}{Model}  

&  SST-2 $\epsilon  = 1$   & SST-2 $\epsilon  = 2$  & SST-2 $\epsilon  = 3$  
& QNLI $\epsilon  = 1$   & QNLI $\epsilon  = 2$  & QNLI $\epsilon  = 3$  \\ \hline
\multirow{1}{*}{Random}   & \multicolumn{3}{|c|} {0.5014} & \multicolumn{3}{|c|}{0.5037}    \\ \cline{2-7} \hline
\multirow{1}{*}{CUSTEXT ~\cite{chen-etal-2023-customized}}  & 0.6985 & 0.7172 & 0.7029 & 0.6926 & 0.6884 & 0.7133   \\ \hline
\multirow{1}{*}{SANTEXT+~\cite{SANTEXT}}  & 0.7211 & 0.7446 & 0.7260  & 0.7607 & 0.7636 & 0.7493         \\ \hline
\multirow{1}{*}{CUSTEXT+~\cite{chen-etal-2023-customized}}   & 0.7501 & 0.7452 & 0.7683 & 0.7528  & 0.7602  & 0.7740 \\ \hline   
\multirow{1}{*}{\textbf{RNM(ours)}}   &  \textbf{0.8265}   & \textbf{0.8331}  &  \textbf{0.8446}   &  \textbf{0.808}   & \textbf{0.8142}  &  \textbf{0.8243}    \\ \hline
\multirow{1}{*}{Original}   & \multicolumn{3}{|c|} {0.9050} & \multicolumn{3}{|c|}{0.9056}    \\ \hline

\end{tabular}
}
}
\label{Table2}
\end{table}

\begin{figure}[t] 
\vspace{-0.3cm}
\centering           
  \subfigure[SST-2] {\includegraphics[width=0.29\textwidth]{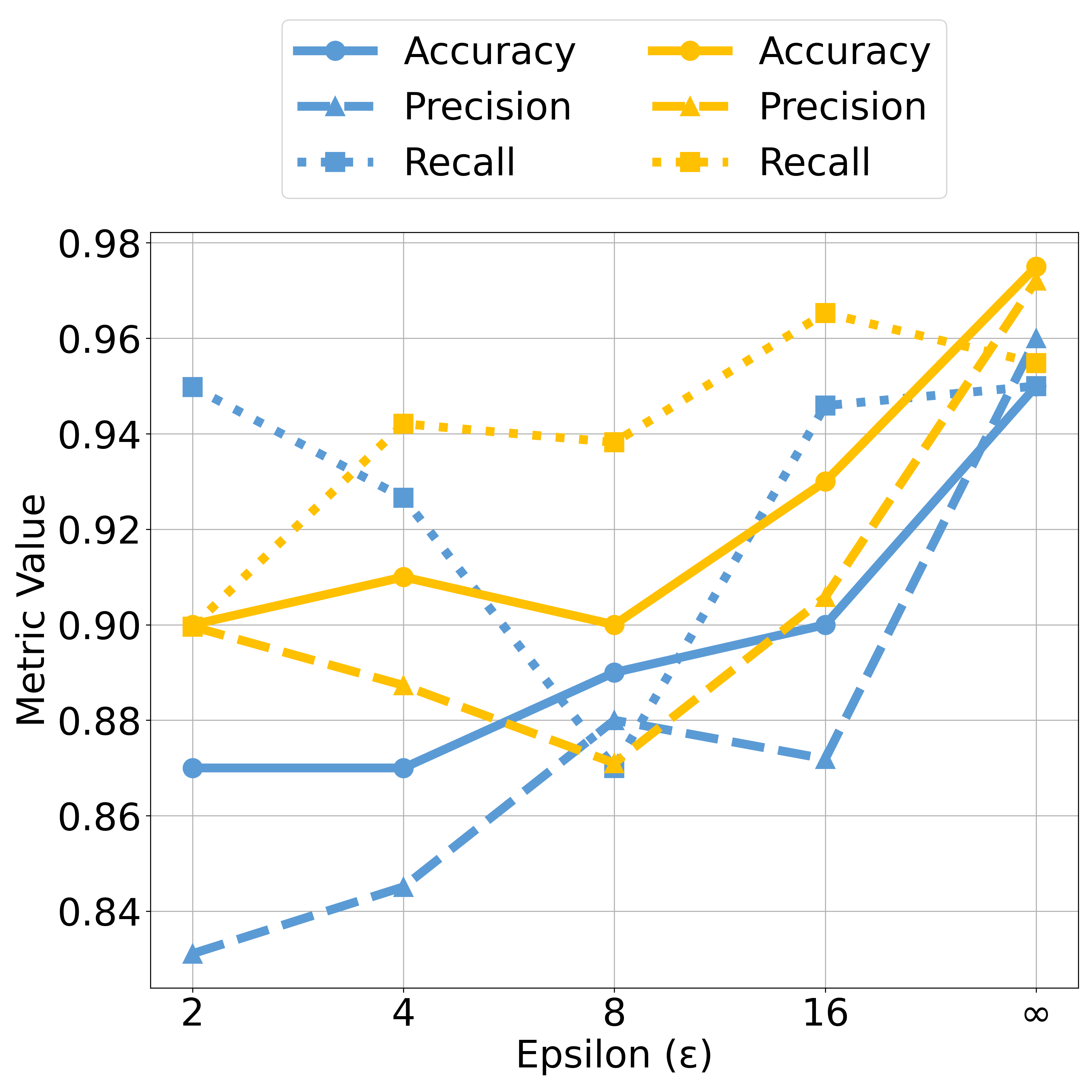}}
  \subfigure[DialogSum] {\includegraphics[width=0.29\textwidth]{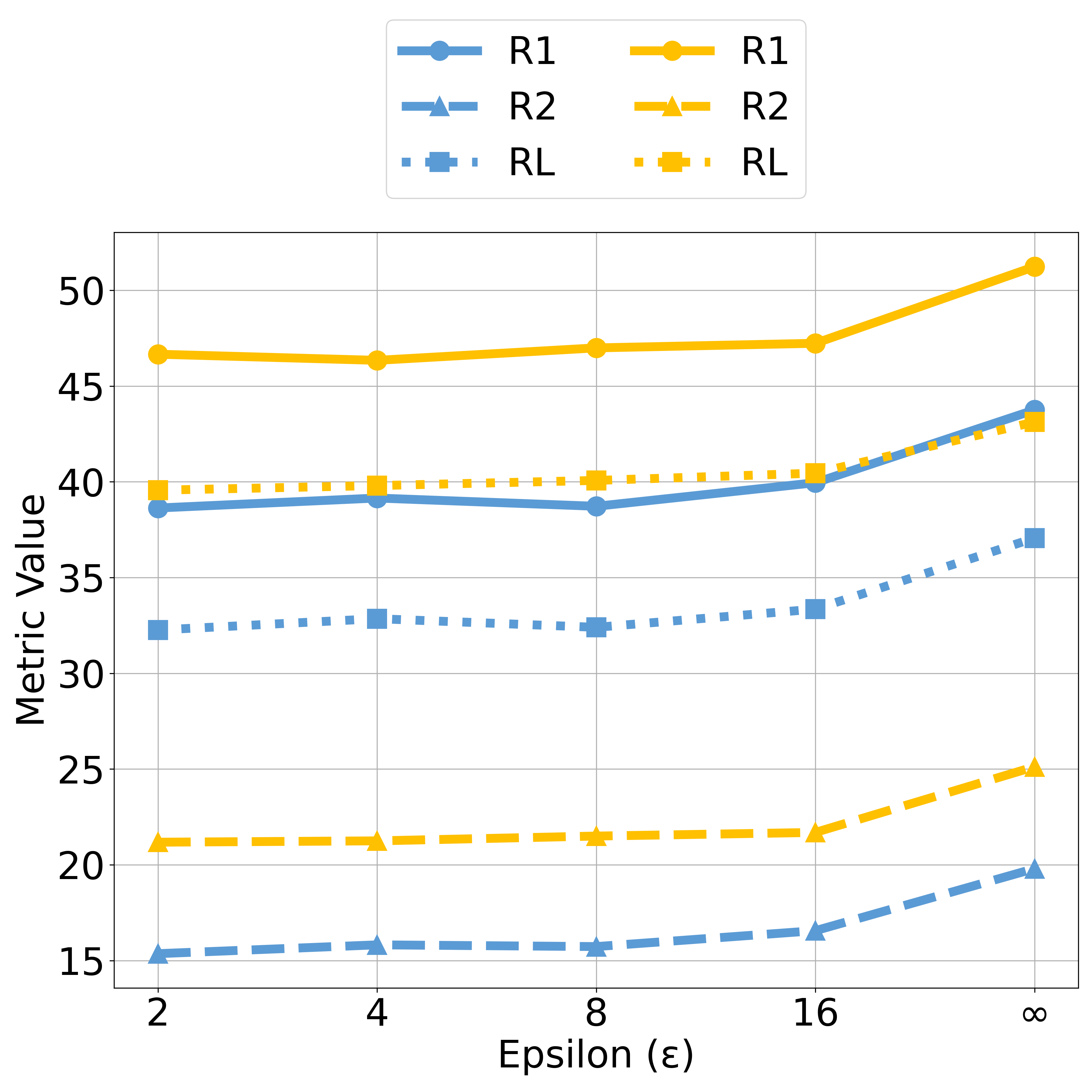}}
   \subfigure[IFEVAL] {\includegraphics[width=0.29\textwidth]{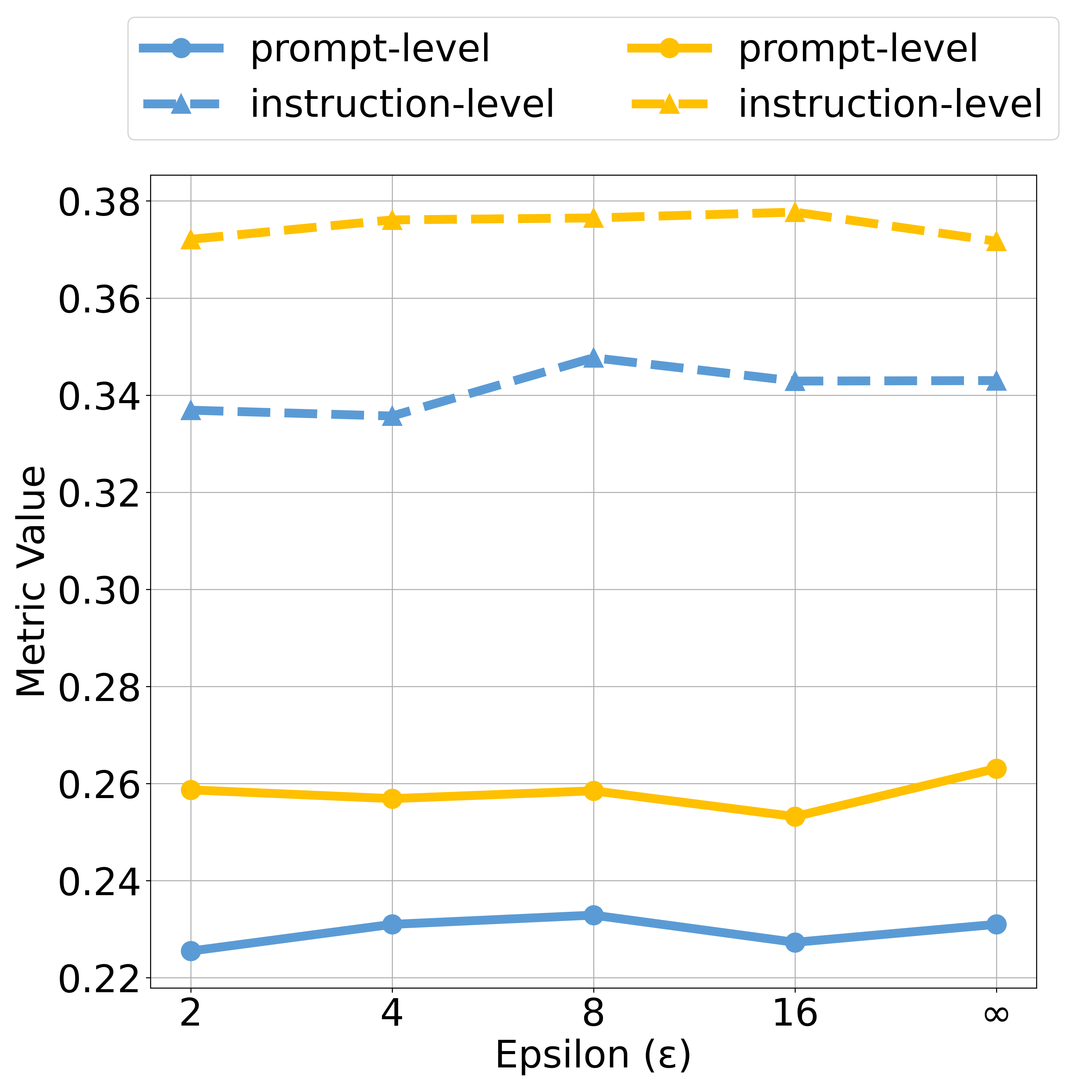}}
\caption{\small Utility comparison of RNM mechanism based on Llama2-7B (the blue curve) and Llama3-8B (the yellow curve)}   
\label{fig3}
\vspace{-0.4cm}
\end{figure}

Subsequently, we conduct experiments on some lightweight LLMs, as shown in Figure~\ref{fig3}.
Llama3-8B outperforms Llama2-7B across three different tasks. It is normal because Llama3-series models have been pre-trained on approximately 15 trillion tokens and have a vocabulary four times larger than Llama2-series models. Consequently, the Llama3-8B model is also more robust when facing noise perturbation. 
On the SST-2 dataset, the model's performance fluctuated. This is because classification tasks are more tolerant of token replacements, so the privacy budget $\epsilon$ is not the dominant role of classification performance.

On the IFEval dataset, we investigate whether fine-tuning on perturbed data will impair the model's other abilities, such as instruction-following. Since the IFEval dataset only provides the prompts for evaluation, we employ the models fine-tuned on DialogSUM. The results demonstrate that the performance of Llama3-8B has declined by 0.15\% compared with when no RNM mechanisms are applied. Nevertheless, both models can successfully complete instructions in the dataset.

\subsection{Inference Efficiency}
As previous TEE-based methods have different prerequisites, it is difficult to make fair comparisons directly. Therefore, we set up two baselines for evaluation.

\textbf{CPU+GPU baseline}: The client executes the embedding layer in a normal environment without input safeguards during communication, while the server processes the remaining model layers on GPUs.

\textbf{GPU baseline}: Entire layers are deployed on GPUs without privacy protection.

\begin{figure}[t] 
\centering
\includegraphics[scale=0.38]{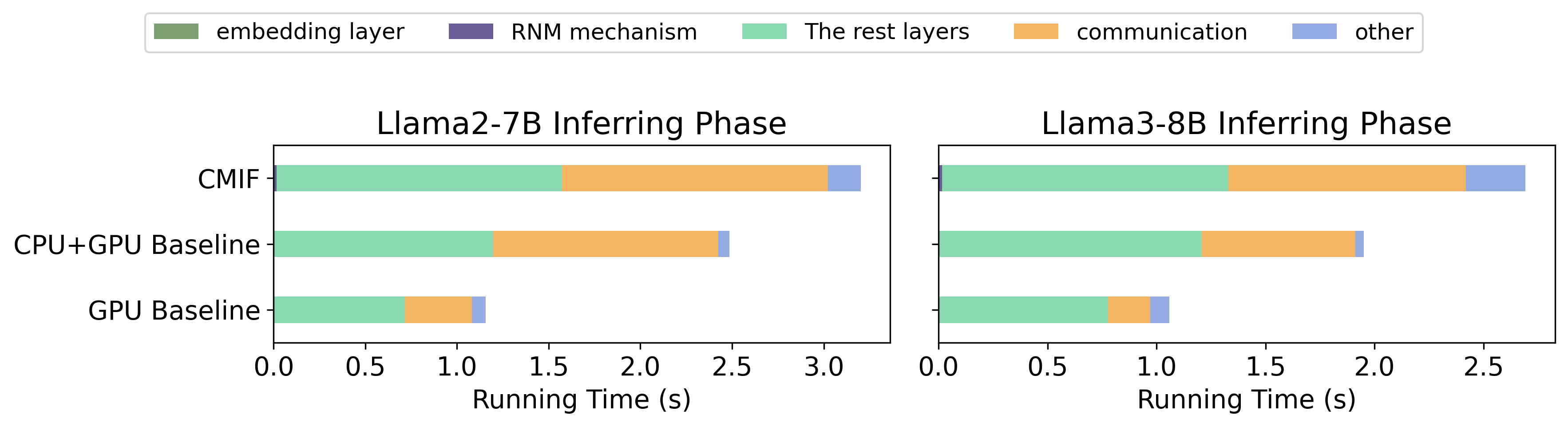} 
\caption{\small Inference time breakdown.}
\label{fig4}
\vspace{-0.4cm}
\end{figure}

As shown in Figure \ref{fig4}, CMIF incurs an additional overhead of 0.7152s (28.78\%) and 0.7432s (38.12\%) for Llama2-7B and Llama3-8B compared with the CPU+GPU baseline. The majority of the cost is attributable to the encryption mechanism of SGX2. The time overhead of the RNM mechanism, including the generation of candidate words and replacements, merely accounts for 0.6\% and 0.74\%. The neighborhood words for each model can be calculated in advance. Generally, CMIF protects the security of user data and model parameters with a low time overhead.
It should be noted that the GPU baseline still exhibits a certain degree of communication overhead, because users must send their data in plaintext to LLMs' servers during inference.
It is also feasible to deploy the embedding layer in an on-device TEE because its parameters account for a minor proportion (about 1.9\% in Llama2-7B) of the entire model and occupy low memory.

\subsection{Data Security}
\label{ds}
In this section, we present a formal security analysis of our sanitization mechanism. 
$x, x'$ are defined as differing in vocabulary $V$. For each $x_i \in V$, its top-k similar words $S_i$ are derived via normalized cosine distance $u$ (sensitivity=1). The RNM mechanism $\mathcal{F}$ replaces $x_i$ with $x^* \in S_i$ by adding Laplace noise $\text{Lap}(1/\epsilon)$ to $u$ and selecting the word with the maximum perturbed score.

\textbf{Case 1 (Both $x, x'$ sensitive):} $\mathcal{F}$ satisfies $\epsilon$-DP (Theorem 1), bounding the probability ratio
\begin{equation}
\Pr(\mathcal{F}(x_i) = \hat{x}) / \Pr(\mathcal{F}(x_i') = \hat{x}) \leq e^\epsilon.
\end{equation}

\textbf{Case 2 ($x$ sensitive, $x'$ non-sensitive):} The bound becomes $e^{\ln(1/p) + \epsilon}$, achieving $(\ln(1/p) + \epsilon)$-DP, where $p$ is the replacement probability for non-sensitive $x'$. A larger $p$ strengthens privacy but degrades utility.

\section{Related Work}

\textbf{TEE-based Methods}
Programs in TEEs are highly secure but incur unacceptable time overhead. Recent studies \cite{ShadowNet,SOTER,Slice2024,huang2024fast} explore to outsource some computations to GPUs for acceleration.
Slalom~\cite{slalom2019iclr} encrypts user inputs $\textbf{x}$ with a
random vector $\textbf{r}$: $f(Enc(X)) = f(X+r)=f(X)+f(r)$. Darknight ~\cite{2021Darknight} uses a simple matrix $\textbf{A}$ to blind $\textbf{X}$ by $f(Enc(\textbf{X}) = f(\textbf{X}\cdot \textbf{A}))$.
Unfortunately, these methods fail to protect the model privacy. SOTER ~\cite{SOTER} protects both models and user inputs by applying a scalar blinding coin $\mu$ to linear operations : $\mu<\textbf{W}\cdot\textbf{X}> = <\mu \textbf{W}\cdot\textbf{X}>$. ShadowNet~\cite{ShadowNet} uses linear transformations to protect linear operators in on-device CNNs. 
Recent studies propose model segmentation strategies for CNNs, such as inserting small slices across layers~\cite{Slice2024} and combining slice strategies with parameter-efficient tuning to protect LLMs' privacy-critical parameters in TEEs ~\cite{huang2024fast}. However, significant overhead remains in TEE-GPU switching and encryption-decryption.

\textbf{DP-based Methods} 
DP-based methods apply anonymization and differential privacy to maintain inference efficiency while ensuring privacy in LLMs. DP-Forward~\cite{DP-Forward} and TextObfuscator~\cite{zhou-etal-2023-textobfuscator} perturb embeddings with calibrated noise, while SANTEXT+~\cite{SANTEXT} uses the exponential mechanism to sample similar tokens. CUSTEXT+~\cite{chen-etal-2023-customized} customizes output tokens to balance privacy and text meaning, but challenges persist in text generation tasks.


\section{Conclusion}
We introduce CMIF, a framework that protects user data through the Report-Noisy-Max sanitization mechanism, eliminating the need for TEE decryption and reducing communication overhead. Additionally, TEEs secure the sanitization process and client-side models. 
Evaluations on diverse datasets have demonstrated the effectiveness of our approach.

\begin{credits}
\subsubsection{\ackname} 
This work was supported by XDB0690302. 

\end{credits}

\bibliographystyle{splncs04}
\bibliography{refs}

\end{document}